\documentclass[9pt,twocolumn,twoside]{opticajnl}
\journal{opticajournal} 

\setboolean{shortarticle}{true}

\usepackage{lineno}

\title{Symmetry-breaking bifurcation of coupled topological edge states}

\author[1,*]{Rujiang Li}
\author[1]{Letian Xu}
\author[1]{Muhammad Imran}
\author[1]{Wencai Wang}
\author[1]{Yongtao Jia}
\author[1]{Ying Liu}

\affil[1]{National Key Laboratory of Radar Detection and Sensing, School of Electronic Engineering, Xidian University, Xi'an, 710071, China}

\affil[*]{rujiangli@xidian.edu.cn}

\begin{abstract}
We propose that the symmetry-breaking bifurcation of coupled topological edge states (CTESs) can be used as a general principle for achieving spontaneous symmetry breaking (SSB) in a nonlinear topological lattice. Using an optical resonator array composed of two Su-Schrieffer-Heeger (SSH) chains as an example, we find that as the nonlinearity strength increases, the symmetric CTESs undergo a supercritical bifurcation. Beyond the critical threshold, the originally stable symmetric state becomes unstable, leading to the formation of a pair of stable asymmetric states. Both sides of the symmetric CTESs exhibit sublattice polarization, while the side of the asymmetric CTESs that is predominantly occupied demonstrates stronger sublattice polarization.
We further find that as interchain coupling increases, the frequency range for stable CTESs expands, while the frequency range for stable asymmetric CTESs decreases.
Our work provides a universal mechanism for realizing SSB in nonlinear topological lattices.
\end{abstract}

\setboolean{displaycopyright}{false} 

\begin{document}

\maketitle

Topological insulators (TIs) are a class of exotic quantum materials that are insulating in their bulk while conducting at their boundaries \cite{hasan2010colloquium,qi2011topological}. The bulk-boundary correspondence, which links the topological invariant of a system to the topologically protected edge states, has laid the foundation for various types of TIs. The concept of TIs and its generalizations have been extended to photonics, leading to the burgeoning field of topological photonics \cite{ozawa2019topological,kim2020recent}. In this domain, light that survives as topological edge states (TESs) exhibits robustness against disorder and defects in complex micro- and nanostructures, highlighting their significant potential for the design of photonic devices \cite{nagulu2022chip,babak_2017,a._2018,zhang__topological_2025}.
In recent years, the investigation of topological photonics has further expanded into the nonlinear regime \cite{daria_2020,alexander_2024}. It is essential to incorporate nonlinear optical effects for TESs at high intensities. Meanwhile, by leveraging intensity-dependent material responses, the intensity of incident light can serve as a dynamic control parameter to manipulate the topological edge states. Extensive studies have been conducted on the formation of nonlinear TESs and solitons \cite{j._2014,yaakov_2016,v._2016,nicolas_2022,daniel_2016,zhaoyang_2020,sebabrata_2021,sebabrata_2020}, nonlinear Thouless pumping \cite{marius_2021,qidong_2022,marius_2023}, topological frequency combs \cite{sunil_2021,j._2024}, etc.

Spontaneous symmetry breaking (SSB), which refers to the sudden collapse of a system's symmetry, is a phenomenon that occurs widely across diverse areas of physics \cite{jeremy_1974}. In nonlinear optical couplers, SSB typically occurs when the nonlinearity strength exceeds a certain critical threshold, leading to the destabilization of an originally stable, symmetric state and resulting in a bifurcation that gives rise to new, stable asymmetric states \cite{m._1989,c._1990,nail_1993,m._1993,a._1996,pengfei_2024}. Recently, it was discovered that edge solitons in a topological trimer array exhibit SSB \cite{y.v._2022}. Specifically, a strongly asymmetric state, with nearly all power concentrated in the edge channel, bifurcates from the in-phase edge solitons with symmetric distributions. However, the realization of such SSB depends on the property of the specific TES. Generally, to achieve SSB in a nonlinear topological lattice in a broader context, it is essential to construct a pair of CTESs and investigate their symmetry-breaking bifurcation.

In this Letter, we propose that a general principle for achieving SSB in a nonlinear topological lattice can be established by utilizing the symmetry-breaking bifurcation of CTESs.
As a specific example, we consider an optical resonator array composed of two SSH chains. We find that as the nonlinearity strength increases, the symmetric CTES undergoes a bifurcation: the originally stable state becomes unstable, leading to the emergence of stable asymmetric states. This bifurcation is classified as supercritical.
Both sides of the symmetric CTESs exhibit sublattice polarization, while the side of the asymmetric CTESs that is predominantly occupied demonstrates stronger sublattice polarization.
In contrast, the antisymmetric CTES does not exhibit any bifurcation. Furthermore, we find that as interchain coupling increases, the frequency range for linearly stable symmetric CTESs expands, while the frequency range for linearly stable asymmetric CTESs decreases. Our work presents a universal mechanism for achieving SSB in nonlinear topological lattices and these results may have applications in the design of nonlinear topological photonic devices.

\begin{figure}[t!]
\centering
\includegraphics[width=8.6cm]{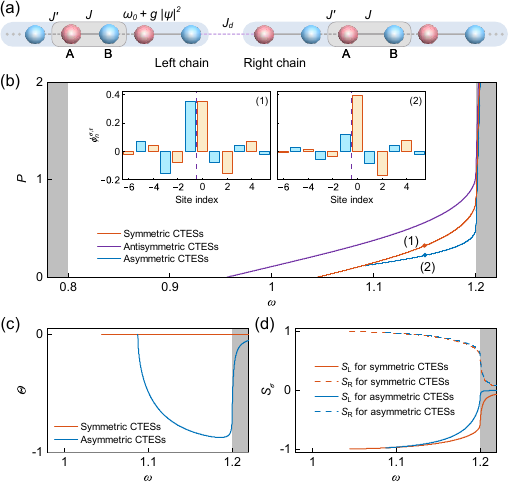}
\caption{Schematic of the resonator array and the symmetry-breaking bifurcation of CTESs.
(a) The resonator array consists of two chains. The resonant frequency of the resonators is expressed as $\omega_{0} + g \left\vert \psi \right\vert^2$, where $\omega_{0}$ is a constant term, $g$ is the Kerr nonlinear coefficient, and $\psi$ represents the amplitude of the optical field at the respective resonator. The intracell and intercell couplings are denoted as $J$ and $J^{\prime}$, respectively, while the two chains are interconnected via a coupling $J_d$.
(b) The frequencies $\omega$ of CTESs at different total powers $P$ are presented. The mode distributions of the symmetric and asymmetric CTESs are shown as insets, with their corresponding frequencies marked on the existence curves. The purple dashed lines in the insets indicate the interface between the two chains.  
(c) The asymmetry parameters $\Theta$ of the symmetric and asymmetric CTESs.  
(d) The sublattice polarizations $s_{\sigma}$ ($\sigma = \mathrm{L}, \mathrm{R}$) of the symmetric and asymmetric CTESs. In panels (b)-(d), the gray regions denote the frequency ranges of the bulk states in the original linear resonator array.}
\label{fig1}
\end{figure}

To illustrate the symmetry-breaking bifurcation of CTESs, we consider an optical resonator array composed of two chains as an example. As shown in Fig.~\ref{fig1}(a), both the left and right chains consist of resonators, with the resonant frequency expressed as $\omega_{0} + g \left\vert \psi \right\vert^2$, where $\omega_{0}$ is a constant term, $g$ is the Kerr nonlinear coefficient, and $\psi$ represents the amplitude of the optical field at the respective resonator. Each unit cell of either chain is composed of resonators A and B, with the intracell and intercell couplings between the resonators denoted as $J$ and $J^{\prime}$, respectively. Additionally, the coupling between the rightmost resonator of the left chain and the leftmost resonator of the right chain is represented by $J_{d}$.
The optical field in the resonator array can be described by the following equations:
\begin{align}
\mathrm{i}\frac{\mathrm{d}}{\mathrm{d}t}\psi_{n}^{\sigma,\mathrm{A}} &= \omega_{0}\psi_{n}^{\sigma,\mathrm{A}} 
+ J \psi_{n}^{\sigma,\mathrm{B}} + J^{\prime}\psi_{n-1}^{\sigma, \mathrm{B}} 
 + \delta_{\sigma, \mathrm{R}} \delta_{n,1} J_{d} \psi_{N}^{\mathrm{L}, \mathrm{B}} \nonumber \\
&\quad + g \left\vert \psi_{n}^{\sigma, \mathrm{A}} \right\vert^2 \psi_{n}^{\sigma, \mathrm{A}}, \label{eq1}\\
\mathrm{i}\frac{\mathrm{d}}{\mathrm{d}t}\psi_{n}^{\sigma, \mathrm{B}} &= \omega_{0}\psi_{n}^{\sigma, \mathrm{B}} 
+ J \psi_{n}^{\sigma, \mathrm{A}} + J^{\prime}\psi_{n+1}^{\sigma, \mathrm{A}}  
 + \delta_{\sigma, \mathrm{L}} \delta_{n,N} J_{d} \psi_{1}^{\mathrm{R},\mathrm{A}} \nonumber \\
&\quad + g \left\vert \psi_{n}^{\sigma, \mathrm{B}} \right\vert^2 \psi_{n}^{\sigma, \mathrm{B}}. \label{eq2}
\end{align}
Here, the index $n = 1, \dots, N$ denotes the cell position within the left or right chain, with $N$ representing the number of unit cells in each chain. The symbol $\sigma$ indicates the chain index, which can be either L (for the left chain) or R (for the right chain). The four Kronecker delta functions $\delta_{\sigma, \mathrm{L}}$, $\delta_{\sigma, \mathrm{R}}$, $\delta_{n,1}$, and $\delta_{n,N}$ characterize the coupling between the left and right resonator chains. Under these definitions, $\psi_{n}^{\sigma,\mathrm{A}}$ and $\psi_{n}^{\sigma,\mathrm{B}}$ represent the mode amplitudes at resonators A and B, respectively, in the $n$th unit cell of the left ($\sigma = \mathrm{L}$) or right chain ($\sigma = \mathrm{R}$). 
Since Eqs. (\ref{eq1})-(\ref{eq2}) are normalized, all physical quantities are dimensionless. The array parameters are set as $\omega_{0} = 1$, $J = 0.1$, $J^{\prime} = 0.3$, $J_{d} = 0.05$, $g = 1$, and $N = 100$. To obtain the stationary solutions, we substitute $\psi_{n}^{\sigma, \tau} \left( t \right) = \phi_{n}^{\sigma, \tau} e^{-\mathrm{i}\omega t}$ ($\tau = \mathrm{A}, \mathrm{B}$) into Eqs. (\ref{eq1})-(\ref{eq2}), where $\omega$ represents the eigenfrequency (the resonant frequency of the entire array) and $\phi_{n}^{\sigma, \tau}$ denotes the eigenmode. The resulting equations can then be solved numerically using Newton's method.

In the weak field limit where $\psi_{n}^{\sigma, \tau} \to 0$, Eqs. (\ref{eq1})-(\ref{eq2}) describe a resonator array that is analogous to two coupled SSH chains \cite{PhysRevLett.42.1698}. When $J < J^{\prime}$, both the left and right resonator chains are topologically nontrivial, supporting TESs \cite{p._2011}. Specifically, when $J_{d} = 0$, the left chain supports a TES localized at its right edge, while the right chain has a TES localized at its left end. Once the interchain coupling $J_{d}$ is introduced, the TESs of the two chains couple to form two CTESs with symmetric and antisymmetric distributions, respectively. Note that the TESs at the ends of the array do not affect the CTESs due to the large lattice size.

We define the total power of all the resonators as $P = P_{\mathrm{L}} + P_{\mathrm{R}}$, where $P_{\sigma} = \sum_{n} \sum_{\tau = \mathrm{A}, \mathrm{B}} \left\vert \phi_{n}^{\sigma, \tau} \right\vert ^{2}$, with $\sigma = \mathrm{L}, \mathrm{R}$ denoting the optical powers in the left and right chains, respectively. As illustrated in Fig.~\ref{fig1}(b), with the increase in the total power $P$,  specifically with the enhancement of nonlinearity strength, both the CTESs with symmetric (red curve) and antisymmetric (purple curve) distributions exhibit a blueshift. When $P$ reaches a critical threshold, the symmetric branch undergoes a bifurcation, leading to the persistence of symmetric CTESs and the emergence of asymmetric CTESs (blue curve). The mode distributions of the symmetric and asymmetric CTESs are shown in the insets of Fig.~\ref{fig1}(b), with their corresponding frequencies $\omega$ marked on the existence curves. Notably, there are two degenerate asymmetric CTESs that are inversion symmetric with respect to the interface between the left and right chains. For clarity, the interface is indicated by purple dashed lines in the insets of Fig.~\ref{fig1}(b). Here, we only present the asymmetric CTES that predominantly resides in the right chain for simplicity. In contrast to the symmetric CTESs, the symmetry-breaking bifurcation of the antisymmetric CTESs is not found (see Supplement 1, Section S1, for details). As $P$ continues to increase, when the frequency enters the range of bulk states in the original linear resonator array (denoted by the gray regions), all the CTESs hybridize with the bulk states and become delocalized.

The symmetry-breaking bifurcation of CTESs can be quantitatively characterized by an asymmetry parameter defined as $\Theta = \left( P_{\mathrm{L}} - P_{\mathrm{R}} \right) / \left( P_{\mathrm{L}} + P_{\mathrm{R}} \right)$. As illustrated in Fig.~\ref{fig1}(c), we have $\Theta = 0$ for the symmetric CTESs and $\Theta \neq 0$ for the asymmetric CTESs. When the frequency enters the range of bulk states in the original linear resonator array, the asymmetry parameter $\Theta$ of the asymmetric CTESs rapidly approaches zero, as the hybridization with the bulk states diminishes the mode asymmetry.

We also investigate the sublattice polarizations defined as 
$S_{\sigma} = \left( P_{\sigma}^{\mathrm{A}} - P_{\sigma}^{\mathrm{B}} \right) / P_{\sigma}$
with $\sigma = \mathrm{L}, \mathrm{R}$ for both symmetric and asymmetric CTESs, where $P_{\sigma}^{\tau} = \sum_{n} \left\vert \phi_{n}^{\sigma, \tau} \right\vert ^{2}$ ($\tau = \mathrm{A}, \mathrm{B}$). The quantities $s_{\mathrm{L}}$ and $s_{\mathrm{R}}$ measure the sublattice polarizations of mode distributions in the left and right chains, respectively.
As shown in Fig.~\ref{fig1}(d), for the symmetric CTESs, near the linear limit, specifically at small frequencies or low total powers, the sublattice polarization $s_{\mathrm{L}}$ approaches $-1$ (solid red curve) and $s_{\mathrm{R}}$ approaches $1$ (dashed red curve). This occurs because the linear TES for a single chain is ideally sublattice polarized, and the symmetric CTESs induced by the interchain coupling $J_{d}$ approximate a linear combination of the two TESs.
As the frequency $\omega$ increases, the sublattice polarizations $s_{\mathrm{L}}$ and $s_{\mathrm{R}}$ of the symmetric CTESs change only slightly, even when the frequency exceeds the critical frequency for symmetry-breaking bifurcation, while the relationship $s_{\mathrm{L}} = -s_{\mathrm{R}}$ holds throughout the process.
In contrast, for the asymmetric CTESs, as indicated by the blue curves, $s_{\mathrm{R}}$ remains nearly equal to that of the symmetric CTES at the same frequency, while $s_{\mathrm{L}}$ deviates from the symmetric CTES, resulting in $s_{\mathrm{L}} \neq -s_{\mathrm{R}}$. This divergence becomes more pronounced at higher frequencies, implying that the side of the asymmetric CTESs that is predominantly occupied demonstrates stronger sublattice polarization. Finally, when the frequency enters the range of bulk states in the original linear resonator array, the sublattice polarizations for both symmetric and asymmetric CTESs quickly decrease to zero, indicating the absence of sublattice polarizations.

We then carry out a linear stability analysis of the symmetric and asymmetric CTESs, while the analysis of the antisymmetric CTESs is presented in Supplement 1. The perturbed solutions are given by 
$\psi _{n}^{\sigma, \tau}(t) =e^{- \mathrm{i}\omega t}\left( \phi
_{n}^{\sigma,\tau}+u_{n}^{\sigma,\tau}e^{- \mathrm{i}\lambda t}+v_{n}^{\sigma,\tau \ast }e^{ \mathrm{i}\lambda ^{\ast}t}\right)$,
where $u_{n}^{\sigma,\tau}$ and $v_{n}^{\sigma,\tau \ast }$ are the eigenmodes of small perturbations, 
and $\lambda \equiv \lambda_{\mathrm{R}} + \mathrm{i} \lambda_{\mathrm{I}}$ is the eigenvalue, with $\lambda_{\mathrm{I}}$ representing the 
growth rate of perturbations. The condition $\mathrm{Max} \left(\lambda_{\mathrm{I}} \right) =0~\left(>0\right)$ indicates that the underlying stationary solutions $\phi_{n}^{\sigma,\tau}$ are linearly stable (unstable). By substituting the perturbed solutions into Eqs. (\ref{eq1})-(\ref{eq2}) and performing linearization, we arrive at an eigenvalue problem which can be numerically solved.

\begin{figure}[t!]
\centering
\includegraphics[width=8.6cm]{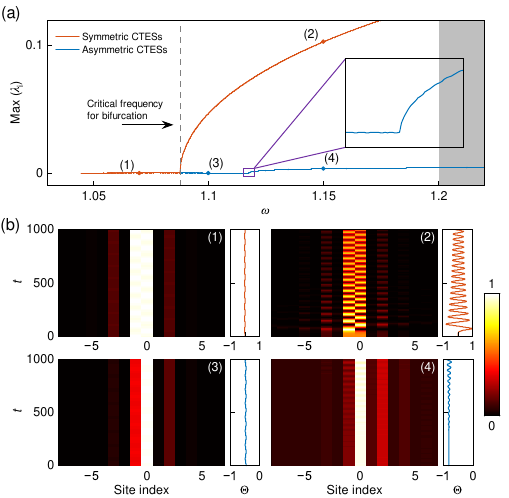}
\caption{Stability analysis of CTESs.
(a) Maximum growth rates, $\mathrm{Max}\left(\lambda_{\mathrm{I}}\right)$, for symmetric and asymmetric CTESs at various frequencies. The critical frequency for symmetry-breaking bifurcation is indicated by the dashed line. 
(b) Temporal evolutions of the symmetric and asymmetric CTESs, with a relative perturbation amplitude of $\pm 2\%$ added to the initial excitations. The time $t$ is dimensionless and state intensities are normalized relative to their maximum values. The frequencies of the CTESs are marked on the curves in panel (a). The time-dependent asymmetry parameter $\Theta(t)$ is plotted alongside each intensity distribution.}
\label{fig2}
\end{figure}

Figure \ref{fig2}(a) shows the maximum growth rates, $\mathrm{Max}\left(\lambda_{\mathrm{I}}\right)$, of the CTESs at various frequencies. The symmetric CTESs are linearly stable before the critical frequency for symmetry-breaking bifurcation and become linearly unstable after bifurcation, as indicated by the solid red curve. This behavior is consistent with that observed in conventional nonlinear optical couplers \cite{m._1989,c._1990,m._1993,a._1996}. However, unlike the asymmetric states that bifurcate from symmetric states in nonlinear couplers, where they are unstable near the bifurcation point and become stable as they move away from it \cite{m._1993,a._1996}, the asymmetric CTESs we have identified are linearly stable only near the bifurcation point but become linearly unstable at higher frequencies, as illustrated by the solid blue curve.

To verify these results, we conduct direct simulations of CTESs at the marked frequencies using Eqs. (\ref{eq1})-(\ref{eq2}), with a relative perturbation amplitude of $\pm 2\%$ added to the initial excitations. The normalized intensity distributions $\left\vert \psi_{n}^{\sigma,\tau} \right\vert^2/ \sum_{n,\sigma,\tau}\left\vert \psi_{n}^{\sigma,\tau} \right\vert^2$ shown in Fig.~\ref{fig2}(b) indicate that the symmetric CTES denoted as state (1) and the asymmetric CTES denoted as state (3) are both dynamically stable, as their temporal evolutions remain nearly stationary. In contrast, the symmetric CTES denoted as state (2) and the asymmetric CTES denoted as state (4) are both dynamically unstable, exhibiting apparent oscillations in their evolutions. To quantitatively evaluate the stationary or oscillatory behaviors, we also plotted the asymmetry parameter $\Theta \left(t\right)$ against dimensionless time $t$ alongside each intensity distribution. It should be noted that the negligible fluctuations for states (1) and (3) are induced by random perturbations, and they remain stable at least up to $t = 10^{5}$.

From the results shown in Figs.~\ref{fig1}-\ref{fig2}, new stable asymmetric CTESs emerge at a power level slightly 
exceeding the threshold where the symmetric CTES becomes unstable. This behavior identifies that the symmetry-breaking bifurcation of symmetric CTESs is a supercritical bifurcation (also known as forward bifurcation or a second-order phase transition) \cite{a._1996, iooss_elementary_1980}. Within the vicinity of this bifurcation point, a unique stable state exists for any given total power $P$, which contrasts with subcritical bifurcation (also referred to as backward bifurcation or first-order phase transition), where bistability typically occurs \cite{a._1996}. The symmetry-breaking bifurcation of CTESs is consistent with the behavior of the symmetric solution of the discrete self-trapping equation, where supercritical bifurcation has been demonstrated \cite{EILBECK1985318}. Furthermore, when the Kerr nonlinear coefficient is negative (i.e., $g < 0$), the symmetry-breaking bifurcation still occurs and still qualifies as a supercritical bifurcation; however, all state branches exhibit negative frequency shifts, and the asymmetric CTESs bifurcate from the antisymmetric branch rather than the symmetric one (see Supplement 1, Section S2).

\begin{figure}[t]
\centering
\includegraphics[width=8.6cm]{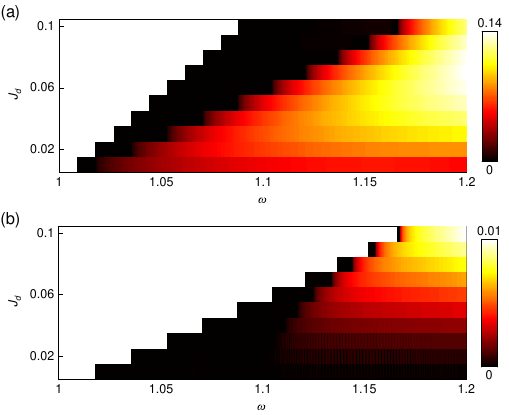}
\caption{Relationship between the frequency ranges of linearly stable CTESs and the interchain coupling $J_{d}$.
(a) Maximum growth rates for perturbed symmetric CTESs at various frequencies and different values of $J_{d}$.  
(b) Same for asymmetric CTESs. The dark regions in panels (a) and (b), corresponding to $\mathrm{Max}\left(\lambda_{\mathrm{I}}\right)=0$, denote the frequency ranges for linearly stable CTESs.}
\label{fig3}
\end{figure}

Finally, we investigate the relationship between the frequency ranges of stable CTESs and the interchain coupling $J_{d}$. Figures \ref{fig3}(a) and (b) present the maximum growth rates for perturbed symmetric and asymmetric CTESs, respectively, across various frequencies and different values of $J_{d}$. 
The symmetry-breaking bifurcation under these parameter values remains a supercritical type.
In Fig.~\ref{fig3}(a), we observe that as $J_{d}$ increases from $0.01$ to $0.1$, the frequency range in which symmetric CTESs exist decreases. This reduction is attributed to the enhanced coupling between the TESs in the left and right chains of the original linear resonator array, which significantly shifts the frequency of the linear symmetric CTES. However, despite the decrease in the existence range, the frequency range for linearly stable symmetric CTESs expands, as indicated by the enlarged dark regions in the figure corresponding to zero maximum growth rates, i.e., $\mathrm{Max}\left(\lambda_{\mathrm{I}}\right)=0$. The increase in the boundary frequency separating linearly stable and unstable states, which is the critical frequency for symmetry-breaking bifurcation, indicates that the frequency range in which asymmetric CTESs exist also decreases, as illustrated in Fig.~\ref{fig3}(b).
In contrast to the behavior observed in symmetric CTESs, the reduced existence range for asymmetric CTESs is accompanied by a decrease in the frequency range for linearly stable asymmetric CTESs, as shown by the shrinking dark regions in Fig.~\ref{fig3}(b). The results depicted in Fig.~\ref{fig3} highlight the necessity of finding an optimal value of $J_{d}$ to simultaneously maximize the frequency ranges for both linearly stable symmetric and linearly stable asymmetric CTESs.

In conclusion, we have investigated the symmetry-breaking bifurcation of CTESs. By examining an optical resonator array composed of two SSH chains, we uncover the supercritical bifurcation of symmetric CTESs, where the originally stable state becomes unstable, leading to the emergence of stable asymmetric states. In contrast to the sublattice polarization observed on both sides of the symmetric CTESs, the side of the asymmetric CTESs that is predominantly occupied exhibits stronger sublattice polarization compared to the other side.
Moreover, we find that as the interchain coupling increases, the frequency range for stable CTESs expands, while the frequency range for stable asymmetric CTESs decreases. Given that the realization principle can be extended to any system supporting CTESs, including waveguide arrays \cite{alexander_2024}, where the time variable $t$ and the resonator frequency $\omega_0$ are respectively mapped to the propagation coordinate and the propagation constant of an individual waveguide, our work presents a universal mechanism for achieving SSB in nonlinear topological lattices.
As a potential extension of our work, it would be pertinent to investigate SSB in more complex lattices, such as two- and three-dimensional TIs and higher-order TIs \cite{ozawa2019topological,kim2020recent}. Our findings may have valuable implications for the design of nonlinear topological photonic devices, such as couplers and switches.

\begin{backmatter}
\bmsection{Funding} National Key Research and Development Program of China (2022YFA1404902); National Natural Science Foundation of China (12104353); Fundamental Research Funds for the Central Universities (QTZX25086); National Natural Science Foundation of China (62271366); 111 Project.

\bmsection{Disclosures} The authors declare no conflicts of interest.

\bmsection{Data availability} Data underlying the results presented in this paper are not publicly available at this time but may be obtained from the authors upon reasonable request.

\bmsection{Supplemental document} See Supplement 1 for supporting content.
\end{backmatter}

\bibliography{sample}

\bibliographyfullrefs{sample}

\end{document}